\documentclass[%
 aip,
 amsmath,amssymb,
 reprint,%
]{revtex4-1}

\usepackage{graphicx}% Include figure files
\usepackage{dcolumn}% Align table columns on decimal point
\usepackage{bm}% bold math
\usepackage{wrapfig}
\usepackage[utf8]{inputenc}
\usepackage[T1]{fontenc}
\usepackage{mathptmx}
\usepackage{etoolbox}

\usepackage{color}

\makeatletter
\def\@email#1#2{%
 \endgroup
 \patchcmd{\titleblock@produce}
  {\frontmatter@RRAPformat}
  {\frontmatter@RRAPformat{\produce@RRAP{*#1\href{mailto:#2}{#2}}}\frontmatter@RRAPformat}
  {}{}
}%
\makeatother
\begin{document}

%\title{High Quality Mesoscopic Circuits on Diamond Anvils}
\title{Sub-micron Circuit Fabrication on Diamond Anvils for Mesoscopic High-Pressure Experiments}
% Force line breaks with \\
\author{Z. R. Rehfuss}
\thanks{These authors contributed equally.}
\affiliation{Department of Physics, Washington University, St. Louis, MO, 63130, USA}
\author{K. Zheng}
\thanks{These authors contributed equally.}
\affiliation{Department of Physics, Washington University, St. Louis, MO, 63130, USA}
\author{S. L. Gould}
\affiliation{Department of Physics, Washington University, St. Louis, MO, 63130, USA}
\author{K. W. Murch}
\affiliation{Department of Physics, Washington University, St. Louis, MO, 63130, USA}
\author{S. Ran}
\email[Email address: ]{rans@wustl.edu}
\affiliation{Department of Physics, Washington University, St. Louis, MO, 63130, USA}

\date{\today}% It is always \today, today,

\begin{abstract}
We present a novel fabrication procedure to produce high-quality lift-off structures on diamond anvils extending from the culet down to the slanted facets. Feature sizes down to 500 nm are achieved through the use of a bi-layer resist stack and electron beam lithography. Device structures with strong adhesion to the diamond surface and high abrasion resistance are realized by optimizing the surface treatment. To benchmark our process, we fabricate a multi-lead tungsten circuit to measure changes of the superconducting transition temperature of zirconium across the structural phase transition at $\sim$30 GPa; showing a 4-fold jump of the critical temperature. Our process is easily reproducible in most traditional academic and industrial cleanroom facilities. This work paves the way for complex and high-precision fabrication and measurements inside diamond anvil cells and on other faceted crystalline samples. 
\end{abstract}

\maketitle
\flushbottom

External Pressure is one of the fundamental thermodynamic variables, intricately related with the thermodynamic equation of state for a given material. By controlling the pressure on the sample, one can selectively tune the properties of the material\cite{MgSiO3, hydrogen, DACReview2D, Qi2018, Bi2022,Yue2023,Anzellini2019, HAMLIN201559, cryst10060459}. Historically, controlling high pressure has lagged behind other thermodynamic variables such as temperature or magnetic field. To overcome the challenge of tuning pressures into the high-pressure regime, diamond anvil cells (DACs) have been utilized in the last century to  push pressure into the terapascal range. 

The DAC was designed in 1959 by Weir et. al.\cite{Weir1959} and was implemented for infrared spectroscopy experiments on calcite and aragonite. This was achieved by clamping powders of interest directly between two diamonds with no gasket or pressure medium. The metal gasket with a hole was introduced by van Valkenburg in 1964\cite{van1964diamond} to increase hydrostaticity. Diamonds were aligned to perfection in 1967 by Bassett et. al. in 1967\cite{Bassett1967}, and the technique was further improved by Piermarini et. al.\cite{Piermarini1975} to reach pressures up to 40 GPa. The ruby pressure scale was tested and calibrated in 1972 by Forman et. al.\cite{Forman1972} by measuring the shift of the Ruby $\rm{R_1}$ and $\rm{R_2}$ lines. Beveled anvils were used by Mao and Bell in 1978\cite{Mao1978} to reach static pressures well above 100 GPa. To date, the highest static pressures generated are Tera-pascal pressures, which were achieved by Dubrovinskaia et. al.\cite{Dubrovinskaia2016} in 2016.

Until the 1970s, all the experimental techniques done inside a DAC were done with optical measurements. Block et. al.\cite{Block1976} devised a two-wire measurement scheme with a gasketed cell and liquid pressure medium. Much work has been done to introduce leads into the diamond anvil cell sample space to perform electrical measurements at high pressures. Sputtered contacts for precision electrodes were introduced in 1984 by Gryzbowski et. al.\cite{Grzybowski1984}, and this paved the way for the modern-day techniques developed for patterned diamond anvils\cite{patterningphotolitho, patdac, Foskaris2022}. Weir et. al.\cite{Weir2000} developed a process for encapsulating the electrode circuits under epitaxially grown diamonds to prevent sheering at high pressures. Boron-doped diamond was also used in some works \cite{dopedDiamond1, dopedDiamond2} as sheer-resistant electrical contact leads. These improved fabrication approaches have advanced electronic characterization at high pressures and prompted a breadth of pioneering studies that deepen our understanding of compressed states of matter \cite{NormNV, GrapheneDAC, MoS2FET}.

%The sputter-deposited leads suffered from one main drawback: they sheer under the stress generated during the experiment. To overcome this problem at high pressure, Weir et. al.\cite{Weir2000} developed a process for encapsulating the electrode circuits under epitaxially grown diamonds. These designer diamonds were shown to be able to withstand very high pressures but suffered from a key limitation of reproducibility for other researchers. The process they developed while high quality, requires complex custom-built equipment for the depositions and violent mechanical polishing after the encapsulation which is not suited for experiments on thin film samples that need to be fabricated on the culet of the diamond. 

%Following the discovery by Wier et. al in 2000, other materials have been used for encapsulation such as Al$_2$O$_3$\cite{Gao2005} which are much easier to deposit than diamond encapsulation layers. This opens up the field of ultrahigh-pressure research to any group that has access to standard clean-room equipment. 

In this work, we demonstrate a high-yield fabrication process for complex mesoscopic circuits on faceted crystalline structures. The process  can be easily reproduced in a typical academic or industrial cleanroom environment. Specifically, we develop a procedure to obtain good focus at the diamond culet for electron beam lithography which allows us to reach feature sizes down to 500~nm with lift-off layers. Strong adhesion is achieved by optimizing the treatment of the diamond surface before metal deposition. 

To benchmark our device performance, we measure the structural phase transition of zirconium metal inside the DAC. We use patterned leads for a four-wire measurement to observe the discontinuous jump in the superconducting transition temperature upon increasing pressure\cite{Zrpaper}. This observation demonstrates the robust performance of the patterned leads in a high-pressure environment.    

% look for a smoking gun signature that can be measured resistively inside our DAC at cryogenic temperatures. Zirconium metal undergoes structural phase transitions upon increasing pressure, correlated with a discontinuous jump in the superconducting transition temperature\cite{Zrpaper}. We use this abrupt change in the 
% $T_c$ as our smoking gun to test the performance of our fabrication process. 

%from the $\alpha$ hexangonal close-packed (HCP) phase at ambient pressure, to the $\omega$ hexagonal phase below 10 GPa, and finally to the high pressure $\beta$ body-centered cubic (BCC) phase around 30 GPa. The superconducting transition temperature ($T_c$) of zirconium metal as a function of pressure shows a monotonic increase from  $\sim$0.6 K at ambient pressure to 4 K at 30 GPa. When Zr is compressed into the high pressure BCC phase, the $T_c$ has a discontinuous jump from 4 K to 11 K\cite{Zrpaper}. We use this abrupt change in the Tc as our smoking gun to test the performance of our fabrication process.

%move this to the conclusion
% The improved device yield and quality, combined with a much smaller critical feature size lays down the foundation of fabricating high quality mesoscopic circuits with more complex designs on diamond culets and faceted crystal samples. 

%Our work lays down the foundation of fabricating high quality mesoscopic circuits on diamond culets, and serves as the first step towards more complex device structures for probing materials properties. 

\begin{figure*}
\centering
    \includegraphics[width=14cm]{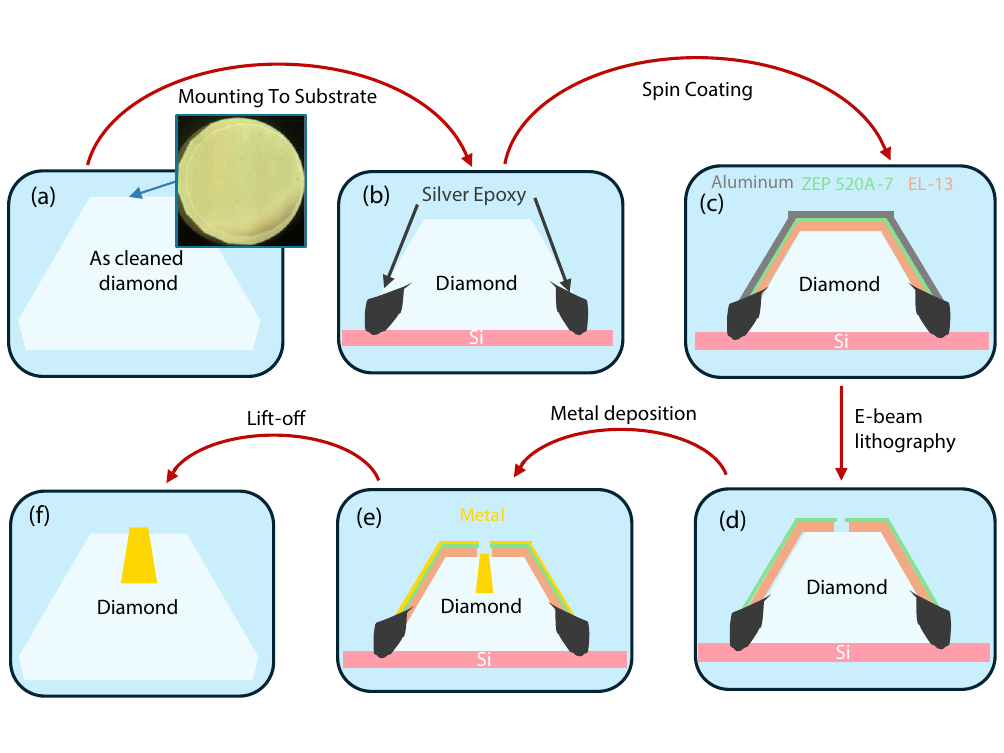}
    \caption{\textbf{Step-by-step  fabrication process to pattern electrical leads on diamond anvils. (a)} A diamond is cleaned following the procedure outlined in the text. (Inset) A picture of the as cleaned culet of a diamond. \textbf{(b)} The cleaned diamond is fixed to a Si carrier chip and attached using silver epoxy. \textbf{(c)} Bilayer electron beam resist is spin coated onto the diamond, the an aluminum discharge layer is deposited on the surface using magnetron sputtering. \textbf{(d)} The leads are patterned using electron beam lithography. \textbf{(e)} Liftoff metal contacts are deposited using physical vapor deposition techniques. \textbf{(f)} A standard lift-off procedure is performed in Kayaku Remover PG leaving behind the patterned diamond.    } 
    \label{processFlow}
\end{figure*}

Figure~\ref{processFlow} provides a step-by-step depiction of our fabrication process to lay down electrical leads. We start with a five step cleaning process. This starts with ultrasonic cleaning in Kayaku Remover PG to remove silver epoxy from previous experiments, then acetone to remove any oils, followed by sequential baths in aqua regia, Transene Nb etchant (HF + HNO\rm{$_3$}), and piranha solution. All cleaning steps are performed for 10~minutes in duration\footnote{Most of the acid treatments are to remove devices fabricated previously on the diamond. For a newly purchased diamond, the acid cleaning procedures may be skipped partially or entirely.}. We secure the as-cleaned diamond onto a 1~\text{\rm{$cm^2$}} carrier chip hand cleaved from a typical Si wafer with silver epoxy (EPO-TEK H20E). We ensure that the diamond table is in direct contact with the chip surface, which ensures that the culet is relatively flat for the lithography step. %A small amount of  silver epoxy is used to secure opposite edges of the diamond table onto the carrier chip before 
We spin coat a bi-layer resist stack of Kayaku PMMA copolymer EL 13 / ZEP 520A-7 electron beam resist onto the diamond surface. When spun at 3000 RPM and soft baked at 180 $\rm{^\circ}$C for 2 minutes, EL 13 has a thickness of 730 nm, and ZEP has a thickness of 270 nm\footnote{We spin the resists separately on a flat Si wafer using the same spin and bake recipe at the same temperature and humidity. The thickness is measured using ellipsometry. We assume that the resist thickness on actual diamond sample is similar to what has been measured on Si wafer.}. We spin 2 layers of EL 13 and a single layer of ZEP in our procedure, yielding a bottom layer thickness of 1460 nm to safely cover the culet-facet sharp edge, and a top layer thickness of 270 nm. The top layer has a higher clear dose than the bottom layer. This allows the bottom layer to form an undercut, which significantly improves the quality of lift-off structures. 

\begin{figure*}
\centering
    \includegraphics[width=14cm]{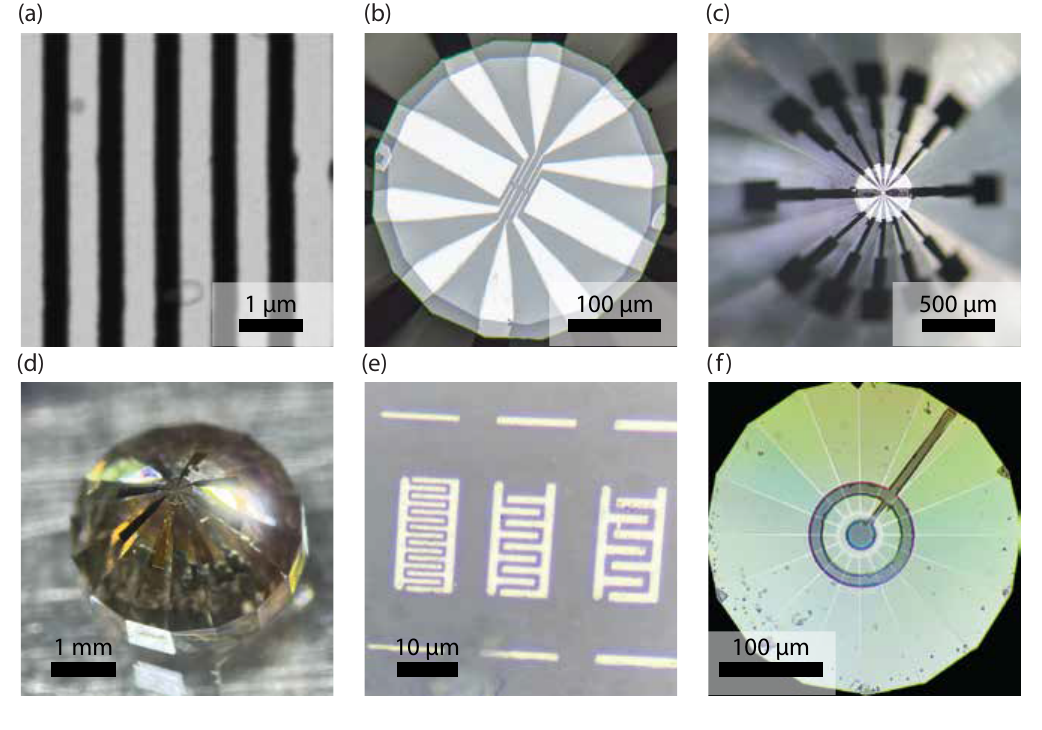}
    \caption{\textbf{Pictures of patterned metal structures on diamond culets.} \textbf{(a)} Scanning electron microscopy (SEM) back-scattering image of  Al/Nb bi-layer features (light color) with 500 nm line width and spacing demonstrates sub-micron resolution . \textbf{(b)} A 10-lead W device with used for the Zr experiment. The metallic islands on the opposing corner of the culet are used to focus the electron beam and to find the center of the culet. \textbf{(c)} A different W device for electrical resistivity measurement. The leads trace continuously from the culet to the facets. \textbf{(d)} A patterned diamond with Au contact pads \textbf{(e)} Nb/Al finger capacitors. \textbf{(f)} A concentric microwave antenna patterned onto 50 \rm{$\mu$}m culet.}
    \label{devices}
\end{figure*}

We then deposit \rm{$\sim$}20 nm of Al on top of the resist inside a Kurt J. Lesker PVD 75 DC magnetron sputter chamber to prevent the pattern from distortion due to charging effects during electron beam lithography. The diamond is then loaded into an Elionix ELS-S50EX lithography system. After focusing the electron beam following the standard procedure of the tool, we search for the diamond girdle vertices. As these features are \rm{$\sim$}1 mm away from the culet, they can be found easily without exposing the culet to the electron beam. The coordinate of each vertex is recorded to construct a regular hexadecagon which allows us to locate the center of the culet within 20 \rm{$\mu$}m accuracy. We jog the beam to one of the corners of the culet and focus the beam at that corner by only moving the Z-height of the sample stage. The view area can be set with a size \rm{$\sim$} 1 \rm{$\mu$}m$^2$ allowing only a negligible fraction of the corner to be exposed during this procedure.  We use two to four opposing culet corners as registration marks to achieve an alignment accuracy better than 100 nm. GenISys BEAMER is used to correct for proximity effects when patterning fine features. After the lithography step, the resist is developed at room temperature in ZED-N50 for 1 minute and in a 1:3 mixture of MIBK:IPA for 2 minutes, followed by a 15-second rinse in IPA.

Before depositing metal, we prepare the diamond surface with a O\rm{$_2$} plasma ashing treatment (100 W power, 15 sccm of O\rm{$_2$} flow, 20~s duration) to promote adhesion. We achieve strong adhesion for Cr, Ti, Ta, Nb, W, Au, Al, and SiO\rm{$_2$} films using either electron beam evaporation or magnetron sputtering. For our prototype device, we deposit 300 nm of W using a Kurt J. Lesker PVD 75 DC magnetron sputter system with base pressure of \rm{$\sim 10^{-7}$} Torr.

After deposition, the diamond is soaked in Remover PG at 65 $\rm{^\circ}$C for 12 hours to promote liftoff of the unwanted metal. The silver epoxy used to attach the diamond onto the carrier wafer softens inside Remover PG and the diamond can then be easily detached from the carrier wafer. The diamond should be free of any silver residue after lift-off, but can be further cleaned by careful abrasion cleaning followed by sonication in acetone if needed. Using the procedures described above, features down to 500 nm can be achieved repeatably as shown in Fig.~ \ref{devices}(a).

To benchmark the capability of our fabrication procedure, a 10-lead device is fabricated for electrical transport measurements using our methods with contact leads of 300 nm W on a single bevel type IIa diamond with a 400/450 \rm{$\mu$}m culet/sub-culet as shown in Fig.~\ref{devices}(b)\footnote{There are two metallic islands on opposing corners of the left and right side of the culet. These are exposed areas used as focusing points and alignment marks during our lithography step. These may be readily removed with a UV laser drill if needed.}. Tungsten leads are chosen due to their resistance to physical abrasion which allows us to repeatedly use the same patterned device to measure multiple samples.

\begin{figure*}
 \begin{center}
    \includegraphics{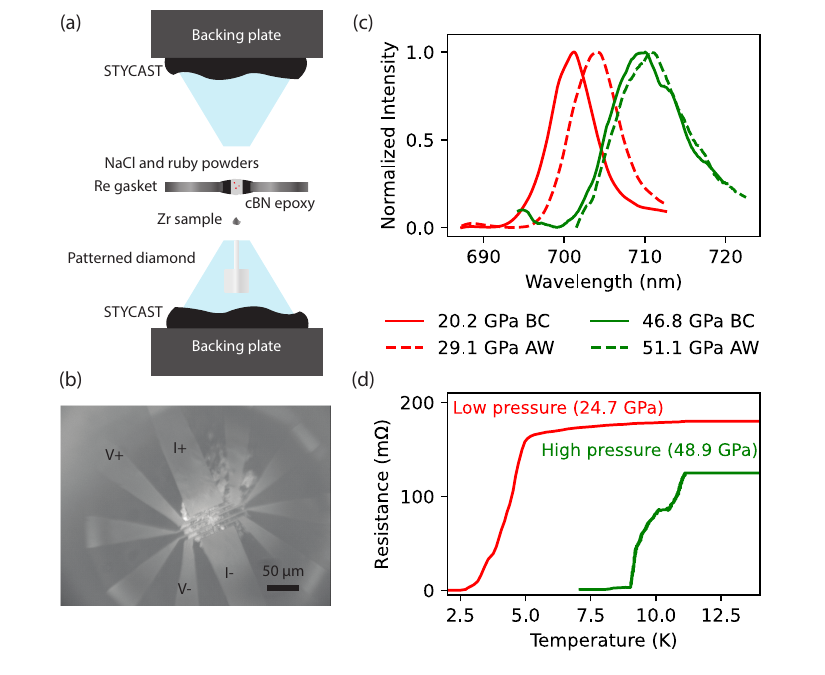}
    \caption{\textbf{Characterization of Zr sample under pressure. (a)} Exploded side view of the loaded DAC to measure the Zr sample. \textbf{(b)} White light image taken through the 10-lead W device patterned diamond with a Zr sample at 51.1 GPa. The 4 wire measurement configuration used for all resistance measurements is shown in black. The 10 leads are clearly seen with no signs of wear up to 51.1 GPa. \textbf{(c)} Ruby florescence collected from a small pile of ruby powder placed in the pressure media directly over the top of the sample. The solid and dashed lines are the spectra before cooling (BC) and after warming up (AW) the clamped cell. The fluorescence profile was used to calculate the pressure\cite{rubyscale}. The average of the pre- and post-cool down pressures was used as the reported pressure.  \textbf{(d)} Temperature dependence of the electrical resistance measured with the 10-lead W device below and above the HCP to BCC structural transition, showing the dramatic enhancement of the critical temperature. The higher pressure (green) data shows signs of pressure inhomogeneity with a clear double transition. This is also reflected in the florescence spectrum with the prominent double shoulder. The inhomogeneity is expected at these high pressures with a solid pressure medium on such a large sample.}
    \label{DAC}
 \end{center}
\end{figure*}

As shown in Fig.~\ref{DAC}(a), a thin sample of Zr with dimensions $\sim 100 \times 50 \times 5$ \rm{$\mu$}m$^3$ is prepared  by polishing a Zr pellet, and placed onto the diamond culet over the patterned circuit using a sharpened toothpick. A set of patterned and anvil diamonds are mounted to 1~cm Inconel backing plates with a 1~mm, 30 degree chamfered through hole using LOCTITE STYCAST 2850FT with LOCTITE CAT 9 adhesive, cured at 65 \rm{$^\circ$}C for 2 hours. The Re gasket is pre-indented to 40 \rm{$\mu$}m and a 350 \rm{$\mu$}m hole is drilled using a drill press. Cubic boron nitride (cBN) nanopowder is mixed with a small amount of STYCAST 1266 A/B and cured at 65 \rm{$^\circ$}C for 2 hours. The resulting cBN epoxy mixture is pressed into and around the drilled hole in the gasket to provide electrical insulation to the leads. To form the sample chamber, a \rm{$\sim$}300 \rm{$\mu$}m hole is cut at the center of the gasket through the epoxy using a \rm{$\sim$}250 \rm{$\mu$}m tungsten carbide end mill. NaCl pressure transmitting medium is loaded into the sample chamber along with fine ruby nano-powder to serve as a manometer\cite{rubyscale}. Finally the cell is closed  and mounted in a screw-driven plate type DAC\cite{Pasternak} to apply pressure to the aligned diamonds. Figure \ref{DAC}(b) shows a optical image of the closed cell at \rm{$\sim$} 15~GPa. 32~gauge Cu wires are silver epoxied to the contact pads on the diamond facets and soldered to external electronics. The pads are patterned to be 500 \rm{$\mu$}m to provide ample space to attach larger gauge wires without the risk of shorting.

%\begin{figure}
% \begin{center}
%    \includegraphics[width=7cm]{fig4_v8.pdf}
%    \caption{\textbf{Characterization of Zr sample under pressure. (a)} White light image taken through the 10-lead W device patterned diamond with a Zr sample at 51.1 GPa. The 4 wire measurement configuration used for all resistance measurements is shown in black. The 10 leads are clearly seen with no signs of wear up to 51.1 GPa. \textbf{(b)} Ruby florescence collected from a small pile of ruby powder placed in the pressure media directly over the top of the sample. The solid and dashed lines are the spectra before cooling (BC) and after warming up (AW) the clamped cell. The fluorescence profile was used to calculate the pressure\cite{rubyscale}. The average of the pre- and post-cool down pressures was used as the reported pressure.  \textbf{(c)} Temperature dependence of the electrical resistance measured with the 10-lead W device below and above the HCP to BCC structural transition, showing the dramatic enhancement of the critical temperature. The higher pressure (green) data shows signs of pressure inhomogeneity with a clear double transition. This is also reflected in the florescence spectrum with the prominent double shoulder. The inhomogeneity is expected at these high pressures with a solid pressure medium on such a large sample.}
%    \label{HighPressure}
% \end{center}
%\end{figure}

We measure the pressure across the culet at room temperature in the initially clamped cell using a 532 nm excitation laser and a SpectraPro HRS-750 spectrometer. The observed ruby fluorescence profile is fit to a Lorentzian distribution to determine the peak location, and thereby determine the pressure\cite{rubyscale}. The initial pressure is determined to be 20.2 GPa as shown in Fig.~\ref{DAC}(c) (red solid line). The cell is mounted into a Bluefors LD250 dilution refrigerator, and a Lake Shore Model 372 AC resistance bridge is used to measure the resistance with 100 \rm{$\mu A$} excitation current in the 632 \rm{$m\Omega$} setting. The resistance vs. temperature at the first clamped pressure is shown in Fig.~\ref{DAC}(d) with a \rm{$T_c$} of 4.3 K. After the measurement, the pressure is measured again at room temperature and is found to have increased to about 29.1 GPa as shown in Fig.~\ref{DAC}(c) (red dashed line). Since we do not posses the ability to measure the pressure at cryogenic temperatures, we take the pressure on the sample to be the average of the pre- and post-cooldown pressures. The pressure is increased incrementally until we achieve a pressure of 46.8 GPa measured by the spectrometer (Fig.~\ref{DAC}(c) (green solid line)). We observe that the superconducting transition is broadened, with two transitions centered at 9.2 K and 10.8 K as seen in Fig.~\ref{DAC}(d). This is due to the pressure inhomogeneity across the sample, which is also reflected in the florescence spectrum Fig.~\ref{DAC}(c) (green solid line), which shows a clear right facing shoulder indicating multiple pressures across the measured area on top of the sample. Figure \ref{DAC}(c) (green dashed line) shows the pressure  to be 51.1 GPa at room temperature after the measurement with a more homogeneous pressure distribution after thermal cycling. 

The dramatic increase in the critical temperature has been previously reported\cite{Zrpaper} due to a first order HCP to BCC structure transition. The agreement between our measured results and published data demonstrates the effectiveness and reliability of our method for patterning diamond anvils. This consistency highlights the precision and robustness of our approach, which successfully achieves micrometer-scale resolution.

We have successfully demonstrated sub-micron lift-off structures fabricated on diamond culets. Using a benchmark device designed for electrical resistivity measurements, we measure the increase of \rm{$T_c$} across a structural phase transition at $\sim$30 GPa, which demonstrates the device's compatibility with high pressure and cryogenic environments. Figure \ref{devices} presents various circuits we have fabricated during process development.  The improved device yield and quality, combined with a much smaller critical feature size lays down the foundation of fabricating high quality mesoscopic circuits with more complex designs on diamond culets and faceted crystal samples. We anticipate our fabrication procedure will enable novel experiments involving, for example, magnetometry using nitrogen vacancy centers in diamond\cite{NormNV}, gated heterostructure and hall bar devices\cite{MoS2FET,GrapheneDAC,DACReview2D}, AC calorimetry using precision deposited thermocouples\cite{ACCT,2WT}, superconducting microwave resonators\cite{Yacoby,Muller_2019}, and AC magnetic susceptibility with counter-wound coils\cite{ACMAG}. We expect our process can be easily reproduced in a typical academic cleanroom. This opens up the door for future groundbreaking experiments using diamond anvil cells.

%\bibliographystyle{unsrt_withcaps}
%\bibliography{sample}% Produces the bibliography via BibTeX.


\section*{Acknowledgments}

The authors would like to thank J. Schilling, C. Zu, and T. Smart for providing the diamond anvil cells and diamonds for this work. The work at Washington University is supported by the National Science Foundation (NSF) Awards DMR-2236528, PHY-2408932, No.~2152221,  and by the Gordon and Betty Moore Foundation, grant DOI 10.37807/gbmf11557.

%\section*{Author Contributions Statement}
%
%Z.R.R. and S.R. conceived the experiments,  K.Z. and Z.R.R. developed the fabrication procedure with support from S.L.G., Z.R.R. conducted the experiments with support from K.Z. and S.L.G., Z.R.R. analyzed the results. S.R. and K.W.M. supervised the work. Z.R.R. and K.Z. wrote the manuscript. All authors reviewed the manuscript. 
%
%\section*{Conflict of Interests}
%
%Z.R.R., K.Z., K.W.M. and S.R. are co-founders of the company Facet Lab. S.L.G. declares no conflict of interests. 
\end{document}